\documentclass[aps,twocolumn,pre,preprintnumbers,amsmath,amssymb]{revtex4-2}
\usepackage[breaklinks=true,colorlinks=true
,urlcolor=blue
,anchorcolor=blue
,citecolor=blue
,filecolor=blue
,linkcolor=blue
,menucolor=blue
,pagecolor=blue
,linktocpage=true
,pdfproducer=medialab
,pdfa=true
]{hyperref}
\usepackage{bbold}
\usepackage{hyperref}
\usepackage[dvipdfmx]{graphicx}
\usepackage{amsmath}
\usepackage{graphicx,float,color}

\definecolor{c1}{RGB}{0,148,17}
\definecolor{c2}{RGB}{255,137,0}

\newcommand{\ket}[1]{| #1 \rangle}
\newcommand{\bra}[1]{\langle #1 |}

\newcommand{\be}{\begin{equation}}
\newcommand{\ba}{\begin{eqnarray}}
\newcommand{\ea}{\end{eqnarray}}
\newcommand{\ee}{\end{equation}}

\newcommand{\bea}{\begin{eqnarray}}
\newcommand{\eea}{\end{eqnarray}}
\newcommand{\bes}{\begin{equation*}}
\newcommand{\beas}{\begin{eqnarray*}}
\newcommand{\eeas}{\end{eqnarray*}}
\newcommand{\bas}{\begin{array*}}
\newcommand{\eas}{\end{array*}}
\newcommand{\ees}{\end{equation*}}

\newcommand{\bpm}{\begin{pmatrix}}
\newcommand{\epm}{\end{pmatrix}}
\newcommand{\bbm}{\begin{bmatrix}}
\newcommand{\ebm}{\end{bmatrix}}

\usepackage{bm}

\begin{document}


\title{
Scrambling Without Chaos in Random Free-Fermionic Systems}
\author{Ali Mollabashi and Mohammad-Javad Vasli
}

\affiliation{
School of Quantum Physics and Matter, Institute for Research in Fundamental Sciences (IPM),
1953833511, Tehran, Iran
}

\begin{abstract}

We study the role of randomness in the scrambling of quantum information within integrable free-fermionic systems. Considering quadratic Hamiltonians with varying degrees of randomness, we analyze entanglement-based measures to characterize the scrambling structure. We show that the memory effect in the entanglement of disjoint subsystems of Gaussian states vanishes when the local couplings are random, indicating information delocalization. The tripartite mutual information exhibits negative saturation values similar to those in chaotic systems, albeit with a smaller magnitude, revealing weaker scrambling under integrable quadratic dynamics. Despite integrability, spectral analyses reveal that local random models display a spectral-form-factor ramp and a partial crossover in the single-particle level-spacing ratio from Poisson-like to Wigner–Dyson-like behavior within a certain range of random couplings. These results demonstrate that randomness can act as a minimal ingredient for inducing information scrambling in integrable quadratic fermionic models.
\end{abstract}
\maketitle
\section{Introduction}
The dynamics of quantum information in many-body systems has emerged as a central theme at the intersection of quantum information, statistical mechanics, and high-energy physics. A key manifestation of this dynamics is information scrambling \cite{Hayden:2007cs, Sekino:2008he, Shenker:2013pqa}, the process by which initially localized information becomes delocalized and effectively inaccessible to local measurements. Scrambling underlies diverse phenomena, from thermalization \cite{PhysRevA.43.2046, PhysRevE.50.888} to the fast scrambling conjecture for black holes \cite{Sekino:2008he, Lashkari:2011yi}. While chaotic systems are typically associated with rapid and efficient scrambling, integrable models exhibit constrained dynamics with highly structured correlations, raising the question of how scrambling unfolds in the absence of chaos. Understanding the interplay between integrability and chaoticity in governing the patterns and efficiency of information scrambling is thus crucial for clarifying fundamental aspects of many-body dynamics.

Information scrambling is characterized by various measures, including the spectral form factor (SFF) \cite{PhysRevE.55.4067, Haake:2010fgh, Cotler:2016fpe}, out-of-time-ordered correlators \cite{Shenker:2013pqa, OTOC, Maldacena:2015waa, Roberts:2014ifa}, tripartite mutual information (TMI) \cite{Hosur:2015ylk, Iyoda:2017pxe, Pappalardi:2018frz}, operator entanglement \cite{Zanardi:2001zza, Alba:2019okd, Bertini:2019gbu, Nie:2018dfe, Styliaris:2020tde}, and entanglement dynamics \cite{Asplund:2015eha, PhysRevX.7.031016, Bertini:2018fbz, Alba:2019ybw}. Recent studies, however, reveal that behaviors such as the exponential growth of out-of-time-ordered correlators \cite{Caputa:2016tgt, PhysRevE.101.010202, Xu:2019lhc, Hashimoto:2020xfr, Trunin:2023xmw, Trunin:2023rwm} and operator entanglement \cite{Dowling:2023hqc} do not necessarily indicate quantum chaos. This prompts the question: What are the minimal conditions for a system to exhibit information scrambling, and can these criteria be established for integrable systems?

In a recent work \cite{Mollabashi:2024gik}, it was shown that a chaotic sector can emerge within a bosonic integrable system and induce information scrambling, despite the system’s global integrability. This sector originates from random quadratic interactions: while the single-particle spectrum exhibits Wigner–Dyson statistics, the full many-body spectrum remains Poissonian. Hence, randomness within an integrable system can act as a driver of scrambling.\footnote{The relation between randomness, scrambling, and complexity is further discussed in \cite{Liu:2017lem}.} The present paper aims to generalize the analysis of \cite{Mollabashi:2024gik}, which focused on bosonic integrable systems, to the fermionic case, and to explore how different degrees of randomness affect information scrambling in such systems. In addition to this generalization from bosons to fermions, the work also provides new insights into the spectral statistics underlying scrambling in these integrable models.

In this paper, we focus on fermionic Gaussian states, which are of particular importance due to their well-established efficient simulability \cite{doi:10.1137/S0097539700377025, PhysRevA.65.032325, Knill:2001lto}. These states have been extensively investigated from various perspectives, including thermalization, eigenstate entanglement structure, and their associated Page curves \cite{Magan:2015yoa, Vidmar_2016, Vidmar:2017uux, Vidmar:2018rqk, Lydzba:2020qfx, Lydzba:2021hml, Bianchi:2021aui, Bianchi:2021lnp, Lydzba:2021zce, Bhattacharjee:2021jff, Lydzba:2022ibr, Yauk:2023wbu, Yu:2022aej}. The aim of this work is to explore information scrambling diagnostics in fermionic Gaussian states undergoing time evolution driven by quadratic fermionic Hamiltonians with varying degrees of randomness. Random quadratic fermionic Hamiltonians are integrable in the sense that, for a system of $N$ fermions, one can identify $N$ conserved operators, namely the occupation number operators of the decoupled modes, that commute with the Hamiltonian.

We study quadratic fermionic Hamiltonians with both local and non-local interactions. In the local models, we analyze the time evolution of the entanglement entropy of disjoint blocks, a well-established scrambling diagnostic \cite{Asplund:2015eha}. Our results show that evolution under an integrable Hamiltonian with sufficient randomness completely eliminates the so-called memory effect, a characteristic feature of integrable systems. We further investigate the dynamics of the tripartite mutual information (TMI) for random local and non-local models, and observe that the TMI becomes negative and saturates at a negative value, closely resembling the behavior found in chaotic systems.

The spectral form factor of SYK$_2$ was analyzed in \cite{Winer:2020mdc, Liao:2020lac}, where the presence of an exponential ramp was reported. In this work, we numerically examine the SFF in both local and non-local models, and similarly observe a non-linear ramp, which appears even in local models with relatively modest randomness.

Motivated by a possible link between level statistics in specific Hilbert-space sectors of integrable models and the behavior of scrambling diagnostics, we investigate the level statistics across our models. We find that, although the single-particle sector of nonlocal models exhibits Wigner-Dyson statistics in its $r$-parameter distribution, local random Hamiltonians behave differently: even when these local models show scrambling features similar to SYK$_2$ (the entanglement memory effect disappears, the TMI saturates to the same negative value found in SYK$_2$, and the spectral form factor develops a ramp), the $r$-parameter distribution of their single-particle sector does not follow Poisson statistics but also fails to display a clear Wigner-Dyson form. Instead, it shows a mixed or intermediate behavior, falling between Poisson and the canonical Wigner-Dyson universality classes.

The rest of this paper is structured as follows: In the next parts of the introduction section we introduce our random models and review our technical tools for this study. In the next section \ref{sec:memory} we study the memory effect in the entanglement of disjoint blocks in local random systems. In Section \ref{sec:TMI} we study the time evolution of tripartite mutual information under local and non-local random Hamiltonians. Section \ref{sec:sff} is devoted to numerical analysis of the spectral form factor, and in Section \ref{sec:level} we analyze the level statistics in our models. 

\subsection{Our Models}
Since our focus is on understanding the impact of randomness on scrambling in integrable systems, our goal is to construct free (integrable) models that incorporate varying degrees of randomness. A moderate level of randomness is introduced by adding disorder with limited strength or range to the couplings of models with local interactions, while the most extreme cases correspond to integrable models featuring completely random, all-to-all couplings.

We consider the most generic quadratic fermionic Hamiltonian defined as
\begin{equation}\label{eq:GSYK}
H=\sum_{i,j=1}^{L}c_{i}^{\dagger}A_{ij}  c_j +\frac{1}{2}\sum_{i,j=1}^{L} \left(c_{i}^{\dagger} B_{ij} c_{j}^{\dagger} + c_{i} B_{ji}^{*} c_{j} \right), \end{equation}
where we will consider different choices for $\mathbf{A}$ and $\mathbf{B}$ matrices. We are interested in both particle-conserving ($\mathbf{B}=0$) and non-particle-conserving ($\mathbf{B}\neq 0$) Hamiltonians. 

A moderate level of randomness is introduced in a disordered version of the Ising model, $H_{\text{Ising}}^{\text{dis.}}$, defined as
\begin{align}
\begin{split}
A_{ij} &= -h\,\delta_{ij} + \frac{1}{2} \big( J_i \, \delta_{j,i+1} + J_j \, \delta_{j,i-1} \big)\,,    
\\
B_{ij} &= J_i \, \delta_{j,i+1} - J_j \, \delta_{i,j+1}\,,
\end{split}
\end{align}
where the parameter $J_i$ is chosen from a random distribution. The case where $J_i$'s are a constant coincides with the fermion representation of the Ising model. Note that in this random model the locality of the system in terms of fermionic operators is preserved while the translational invariance is broken with the random interactions at different positions. We refer to this model as the local random model. 

We also consider fully random models. The case with $\mathbf{A}$ drawn from the Gaussian Unitary Ensemble (GUE) and $\mathbf{B}=0$ is referred to as the Dirac SYK$_2$ model \cite{Lydzba:2021hml}. When both $\mathbf{A}$ and $\mathbf{B}$ are non-vanishing random matrices, we will refer to the resulting system as the generalized SYK$_2$ (GSYK$_2$) model.

These random quadratic models can be diagonalized numerically, and the corresponding procedure will be reviewed in the following subsections. Moreover, since our focus is on entanglement-based information-scrambling diagnostics in fermionic Gaussian states, fully characterized by their correlation matrices, we also review the computation of the correlation matrix below.

\subsection{Diagonalization}
Here we review the diagonalization of a general fermionic quadratic Hamiltonian given in Eq. \eqref{eq:GSYK}. 
In This Hamiltonian $\mathbf{A}$ is a Hermitian matrix ($\mathbf{A} = \mathbf{A}^{\dagger}$) and $\mathbf{B}$ is an antisymmetric matrix (\( \mathbf{B} = -\mathbf{B}^{T} \)). We rewrite the Hamiltonian in a compact form:
\begin{equation}
    H = \frac{1}{2} \Psi^{\dagger} \mathbf{M} \Psi, \qquad
    \mathbf{M} = \begin{pmatrix}
        \mathbf{A} & \mathbf{B} \\
        -\mathbf{B}^{*} & -\mathbf{A}^{T}
    \end{pmatrix},
\end{equation}
where the Nambu spinor \( \Psi \) combines the fermionic creation and annihilation operators:
\begin{equation}
    \Psi^{\dagger} = \begin{pmatrix}
        c_1^{\dagger} & c_2^{\dagger} & \dots & c_L^{\dagger} & c_1 & c_2 & \dots & c_L
    \end{pmatrix}
    \equiv
    \begin{pmatrix}
        \mathbf{c}^{\dagger} & \mathbf{c}
    \end{pmatrix}.
\end{equation}

To diagonalize the Hamiltonian, we consider a unitary transformation \( \mathbf{U} \), constructed from the eigenvectors of the matrix \( \mathbf{M} \), such that:
\begin{equation}
    H = \frac{1}{2} \Psi^{\dagger}  \mathbf{M}  \Psi
    = \frac{1}{2}
    \begin{pmatrix}
        \boldsymbol{\eta}^{\dagger} & \boldsymbol{\eta}
    \end{pmatrix}
    \begin{pmatrix}
        \mathbf{\Lambda} & 0 \\
        0 & -\mathbf{\Lambda}
    \end{pmatrix}
    \begin{pmatrix}
        \boldsymbol{\eta} \\
        \boldsymbol{\eta}^{\dagger}
    \end{pmatrix},
\end{equation}
where the new quasiparticle operators are defined as:
\begin{equation}\label{quasiparticle}
    \begin{pmatrix}
        \boldsymbol{\eta} \\
        \boldsymbol{\eta}^{\dagger}
    \end{pmatrix}
    = \mathbf{U}
    \begin{pmatrix}
        \mathbf{c} \\
        \mathbf{c}^{\dagger}
    \end{pmatrix}.
\end{equation}

The unitary matrix \( \mathbf{U} \) can be decomposed into four blocks:
\begin{equation}
    \mathbf{U} =
    \begin{pmatrix}
        \mathbf{g} & \mathbf{h} \\
        \mathbf{h}^{*} & \mathbf{g}^{*}
    \end{pmatrix}, \qquad
    \mathbf{U}^{\dagger} =
    \begin{pmatrix}
        \mathbf{g}^{\dagger} & \mathbf{h}^{T} \\
        \mathbf{h}^{\dagger} & \mathbf{g}^{T}
    \end{pmatrix}.
\end{equation}

This leads to the Bogoliubov transformation between the original fermionic operators and the quasiparticles:
\begin{equation}
    \mathbf{c} = \mathbf{g}^{\dagger} \boldsymbol{\eta} + \mathbf{h}^{T} \boldsymbol{\eta}^{\dagger}, \qquad
    \mathbf{c}^{\dagger} = \mathbf{h}^{\dagger} \boldsymbol{\eta} + \mathbf{g}^{T} \boldsymbol{\eta}^{\dagger}.
\end{equation}

In component form, this becomes:
\begin{equation}
    c_k = \sum_{i=1}^{L} \left( g^{*}_{ki} \eta_i + h_{ki} \eta_i^{\dagger} \right), \qquad
    c_k^{\dagger} = \sum_{i=1}^{L} \left( h^{*}_{ki} \eta_i + g_{ki} \eta_i^{\dagger} \right).
\end{equation}

Using these quasiparticle operators, the Hamiltonian takes the diagonal form:
\begin{equation}
    H = \sum_{k=1}^{L} \lambda_k \eta_k^{\dagger} \eta_k + \text{const},
\end{equation}
where \( \lambda_k \) are the eigenvalues of the matrix \( \mathbf{M} \), corresponding to the single-particle excitations of the system. We can also define the generalized correlation matrix for fermions in the Nambu representation
		
			\begin{equation}
    \mathcal{C}= \langle \Psi \Psi^{\dagger} \rangle=\begin{pmatrix}
      \langle \mathbf{c} \mathbf{c}^{\dagger} \rangle   & \langle \mathbf{c} \mathbf{c} \rangle \\
       \langle \mathbf{c}^{\dagger} \mathbf{c}^{\dagger} \rangle & \langle \mathbf{c}^{\dagger} \mathbf{c} \rangle
    \end{pmatrix}=\begin{pmatrix}
      \mathbf{g}^{\dagger} \mathbf{g}^{T}   & \mathbf{g}^{\dagger} \mathbf{h}^{T}  \\
       \mathbf{h}^{\dagger} \mathbf{g}^{T}  &  \mathbf{h}^{\dagger} \mathbf{h}^{T}
    \end{pmatrix}
\end{equation}

\subsection{Review of Correlator Method}
	In the following, we employ a method based on a simple yet powerful idea, the so-called correlator method \cite{Peschel_2003, PhysRevLett.90.227902, Latorre:2003kg, Casini:2009sr}.  
	The starting point is a fermionic many-body system in its ground state.  
	For a given spatial subregion $A$, we define the reduced density matrix as
	\[
	\rho_A = \mathrm{Tr}_{\bar A}\, |\mathrm{GS}\rangle\langle \mathrm{GS}|,
	\]  
	where $\bar A$ denotes the complement of $A$.
	
	\medskip
	By construction, $\rho_A$ reproduces the correct expectation values of all operators localized in $A$:
	\[
	\langle O_A \rangle = \mathrm{Tr}(\rho_A \, O_A), \qquad \forall \, O_A \text{ supported in } A.
	\]  
	Thus, $\rho_A$ is the unique object that encodes all local correlators within $A$. Equivalently, full knowledge of all local correlators in $A$ is sufficient, in principle, to reconstruct $\rho_A$ completely.
		
	\medskip
	In the Nambu representation, it is convenient to introduce the generalized correlation matrix restricted to $A$:
	\begin{equation}
		\mathcal{C}_A \equiv \langle \Psi_A \Psi^{\dagger}_A \rangle
		=\begin{pmatrix}
			\langle \mathbf{c}_A \mathbf{c}_{A}^{\dagger} \rangle   & \langle \mathbf{c}_A \mathbf{c}_A \rangle \\
			\langle \mathbf{c}_{A}^{\dagger} \mathbf{c}_{A}^{\dagger} \rangle & \langle \mathbf{c}_{A}^{\dagger} \mathbf{c}_A \rangle
		\end{pmatrix},
	\end{equation}
	where $\mathbf{c}_A$ denotes the vector of annihilation operators in region $A$, and $\Psi_A$ the corresponding Nambu spinor.

    \begin{figure*}[t]
\begin{center}
\includegraphics[scale=0.4]{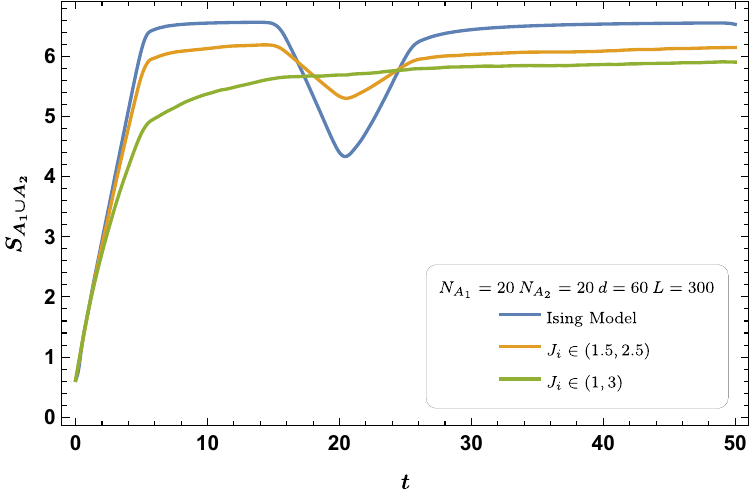}
\hspace{1mm}
\includegraphics[scale=0.4]{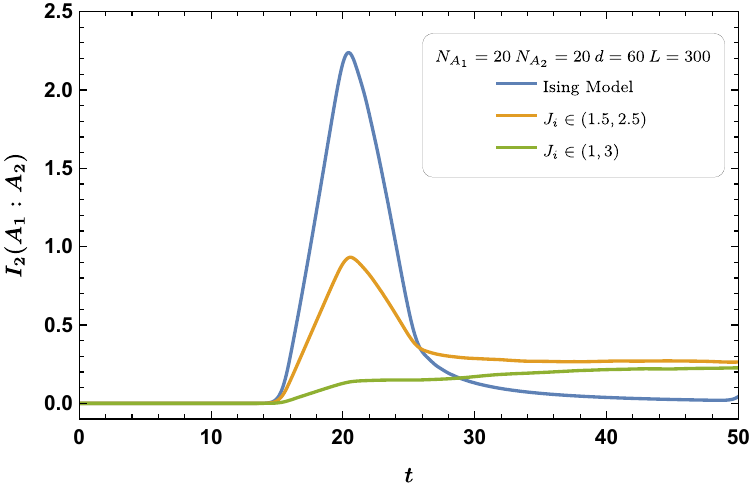}
\hspace{1mm}
\includegraphics[scale=0.4]{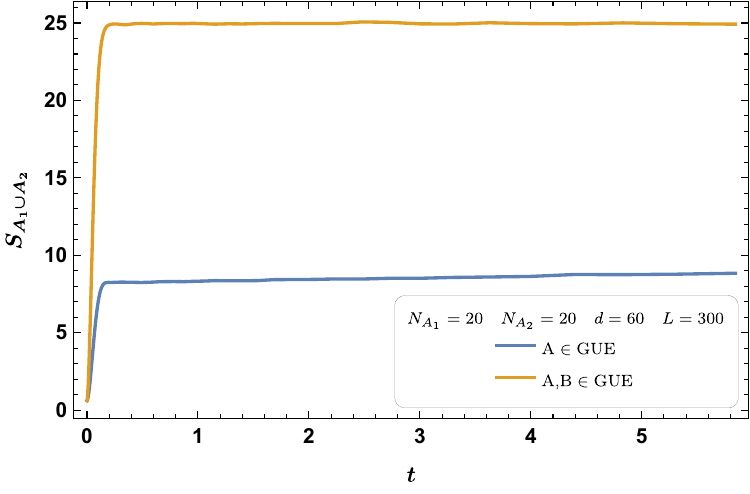}
\end{center}
\caption{\textit{Left:} Time evolution of $S_{A_1\cup A_2}$ in Ising model, local disordered Ising model, and non-local models. The subregion configuration in specified in the legend where $L$ denotes the size of the total system. $J_i$ is chosen from uniform random distribution specified in the plot legend. \textit{Middle:} Time evolution of mutual information corresponding to the left panel. \textit{Right:} Time evolution of $S_{A_1\cup A_2}$ in non-local random models. In all plots corresponding to random models, the results are averaged over 20 samples.
}
\label{fig:SAB}
\end{figure*}

	\medskip
	On the other hand, general arguments imply that the reduced density matrix must take a Gaussian form \cite{Peschel_2003},
	\begin{equation}
		\rho_A = \frac{1}{Z}\,
		\exp\!\Big( -\tfrac{1}{2}\Psi^\dagger_A M_A \Psi_A \Big),
	\end{equation}
	where 
	$
	\mathcal{H}_A = \tfrac{1}{2}\Psi^\dagger_A M_A \Psi_A
	$
	is the so-called \emph{modular Hamiltonian} (or entanglement Hamiltonian) in Nambu space, and $Z$ ensures normalization.  
	Since $\rho_A$ is Hermitian, the matrix $M_A$ must itself be Hermitian, and therefore it can be diagonalized via a suitable Bogoliubov transformation (see Eq.~\eqref{quasiparticle}).  
	As a result, $\rho_A$ can be brought into the diagonal form
	\begin{equation}
		\rho _{A}=\prod_l \frac{e^{-\epsilon _{l}\eta_{l}^{\dagger }\eta_{l}}}
		{1+e^{-\epsilon _{l}}}, \label{diago}
	\end{equation}
	where $\{\eta_l\}$ denote quasiparticle operators and $\epsilon_l$ are the single-particle energies.  
	Also, we use normalization factor 
	$
	Z=\prod_l \big(1+e^{-\epsilon_l}\big).
	$

	\medskip
	Next, we establish the precise relation between the generalized correlation matrix $\mathcal{C}_A$ and the modular Hamiltonian $\mathcal{H}_A$.  
	By definition,
	\begin{equation}
		\mathcal{C}_A = \mathrm{Tr}(\rho_A \, \Psi_A \Psi_A^\dagger).
	\end{equation}
	Evaluating this using the diagonal form \eqref{diago}, one finds that the eigenvalues $\epsilon_l$ of $\mathcal{H}_A$ are related to the eigenvalues $\nu_l$ of $\mathcal{C}_A$ by
	\begin{equation}
		\nu_l = \frac{1}{1+e^{\epsilon_l}}
		\qquad \Longleftrightarrow \qquad
		e^{\epsilon_l} = \frac{1-\nu_l}{\nu_l},
	\end{equation}
	with $\nu_l \in (0,1)$ \cite{Peschel_2003}.
	
	\medskip
	Finally, the entanglement entropy, defined as the von Neumann entropy of $\rho_A$,
	$
	S_A=-\mathrm{Tr}\!\left(\rho_A\log\rho_A\right),
	$
	admits two equivalent and practically useful representations:
	\begin{align}\label{EE}
		S(A) &= \sum_l \left[ \log \!\left(1+e^{-\epsilon _{l}}\right)
		+ \frac{\epsilon _{l}\, e^{-\epsilon _{l}}}{1+e^{-\epsilon _{l}}}\right], \\[6pt]
		&= -\sum_{l}\Big[ (1-\nu_l)\log(1-\nu_l)+\nu_l \log \nu_l \Big].
	\end{align}

\section{Memory Effect}\label{sec:memory}
In a local system, the time evolution of entanglement between two disjoint intervals and the complement region is a well known diagnostic for information scrambling. Here we denote the disjoint intervals by $A_1$ and $A_2$ and their separation by $d$. For the case where $\ell_{A_1}=\ell_{A_2}$ that we chose for simplicity, when $d>\ell_{A_1}$, $S_{A_1\cup A_2}(t)$ shows a memory effect in integrable systems, where no global scrambling is expected, and such an effect does not appear in chaotic systems \cite{Asplund:2015eha}. The memory effect manifests as a dip-ramp pattern in the saturation of $S_{A_1 \cup A_2}(t)$, where $A_1$ and $A_2$ become correlated during the dip and uncorrelated during the ramp. This behavior, observed in various studies \cite{Asplund:2015eha, Fagotti:2010yr, Leichenauer:2015xra, PhysRevX.7.031016, Foligno:2024ymi}, is reflected in the mutual information between $A_1$ and $A_2$ as the presence or absence of a spike. It can be physically interpreted through the ballistic propagation of quasi-particles in integrable systems \cite{Calabrese:2005in, Alba:2017ekd, Alba:2018hie}.

In Fig.~\ref{fig:SAB} we present numerical simulations of the quench dynamics. Throughout our analysis, the initial (pre-quench) state is chosen as the ground state of the Ising Hamiltonian with the transverse magnetic field shifted by one unit, i.e. $h_c+1$, where $h_c$ denotes the critical value of $h$. This state is then evolved under the random Hamiltonian $H_{\text{Ising}}^{\text{dis.}}$, and the time dependence of the entanglement entropy is obtained by numerical diagonalisation of $\mathcal{C}$. As shown by the blue curve, the clean Ising model exhibits a clear memory effect. Introducing randomness in the couplings $J_i$ progressively suppresses this effect: it becomes shallower for $J_i \in (1.5,2.5)$ and disappears entirely when $J_i$ is sampled from $J_i \in (1,3)$ or from broader windows centered at $2$. The mutual information panel illustrates the same phenomenon from a complementary perspective. The peak height of the mutual information decreases with increasing randomness, as demonstrated by the green and red curves, which show how randomness \textit{delocalizes} information and diminishes the mutual-information peak. 

The entanglement dynamics of these integrable systems is generally expected to follow a quasi-particle picture \cite{Calabrese:2005in, Alba:2017ekd}. However, as in the bosonic case discussed in \cite{Mollabashi:2024gik}, for random Hamiltonians such as $H_{\text{Ising}}^{\text{dis.}}$ the dispersion relation in terms of rapidities is not well defined. Consequently, applying a predictive quasi-particle picture is questionable. Nevertheless, a heuristic quasi-particle description, where rapidities and entropy densities are sampled from a random distribution, can still capture the qualitative features of the evolution. We do not pursue such an analysis here and instead refer the reader to \cite{Mollabashi:2024gik}, where this approach was carried out explicitly.

For sort of completeness, in the right panel of Fig.~\ref{fig:SAB}, we present our results for the memory effect in systems with all-to-all random interactions namely (G)SYK$_2$. As expected, no memory effect is observed in these cases. The plots indicate that non–particle-conserving random Hamiltonians yield a higher saturation entropy, which is physically reasonable.

\section{Tripartite Mutual Information}\label{sec:TMI}
Another measure for scrambling is the tripartite mutual information defined as
\begin{align}\label{eq:TMI}
\begin{split}    
I_3(A_1&:A_2:A_3)=
\\
&I(A_1:A_2)+I(A_1:A_3)-I(A_1:A_2\cup A_3)\;.
\end{split}
\end{align}

TMI does not possess a definite sign for generic states in extended quantum systems, including those described by quantum field theory \cite{Casini:2008wt, Rangamani:2015qwa, Rota:2015wge, Agon:2021lus}. In contrast, for holographic states, TMI is known to be always negative \cite{Hayden:2011ag}.

Expressed in terms of mutual information, TMI serves as a useful tool for characterizing scrambling. The authors of \cite{Hosur:2015ylk} demonstrated that when TMI attains and saturates at relatively large negative values, it strongly indicates scrambling. This behavior arises due to the disparity between the information gained from measurements on individual subsystems (the first two terms on the right-hand side of Eq. \eqref{eq:TMI}) and that obtained from joint measurements (the last term of the right-hand side of Eq. \eqref{eq:TMI}). Specifically, when the information accessible via measurements on $A_2\cup A_3$ significantly exceeds that gained separately from $A_2$ and $A_3$, TMI takes on large negative values. The most negative value is reached for systems evolved under dynamics induced by a random Haar unitary \cite{Hosur:2015ylk}.

In free integrable systems, the dynamics of TMI have been investigated \cite{Carollo:2022lrq, Maric:2022rsc, Parez:2022egh, Maric:2020dpw, Caceffo:2023hns, Shukla:2023xbt, Berthiere:2024sio}, where negative values at intermediate times have been reported in certain cases \cite{Caceffo:2023hns, Shukla:2023xbt}, and even a saturation to a negative steady-state value has been observed as a result of continuous monitoring \cite{Carollo:2022lrq}.

\begin{figure}[h]
\begin{center}
\includegraphics[scale=0.45]{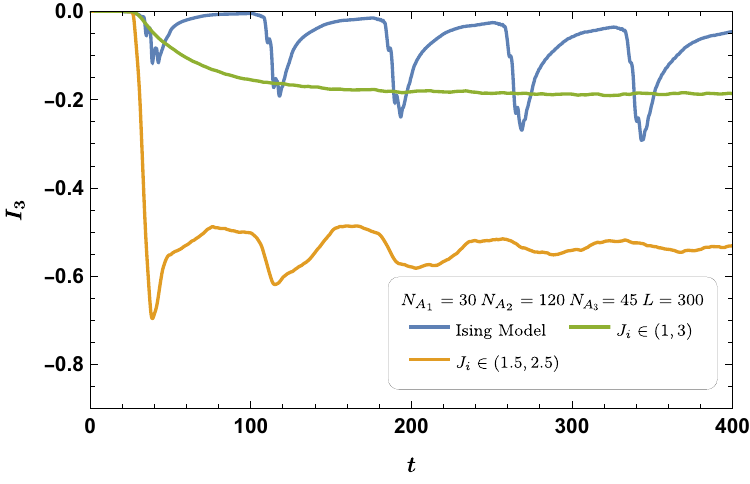}
\hspace{1mm}
\includegraphics[scale=0.45]{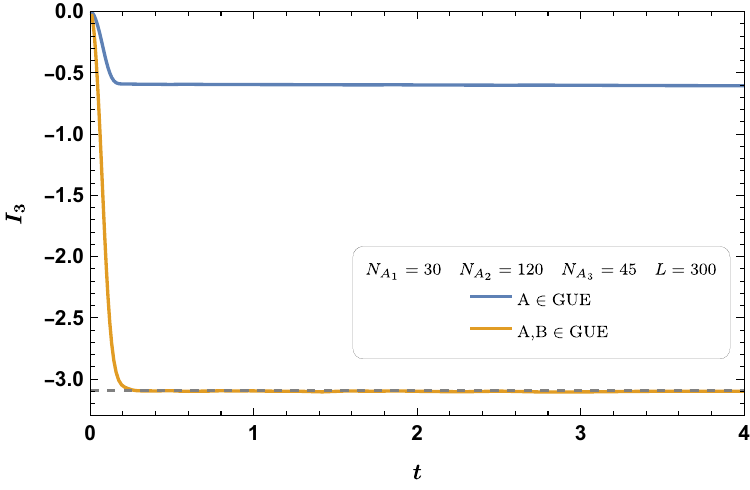}
\end{center}
\caption{\textit{Upper:} Time evolution of Tripartite Mutual Information in Ising model versus local disordered Ising model. The configuration is specified in the legend. $J_i$ is chosen from a uniform random distribution specified in the plot legend. \textit{Lower:} Time evolution of Tripartite Mutual Information in non-local random models. The dashed line indicates the lower bound of the TMI for Gaussian states evolving under quadratic Hamiltonians. In all plots, the results are averaged over 20 samples, except for the Ising model, which does not contain any random parameters.
}
\label{fig:I3}
\end{figure}

\begin{figure*}[t]
\begin{center}
\includegraphics[scale=0.4]{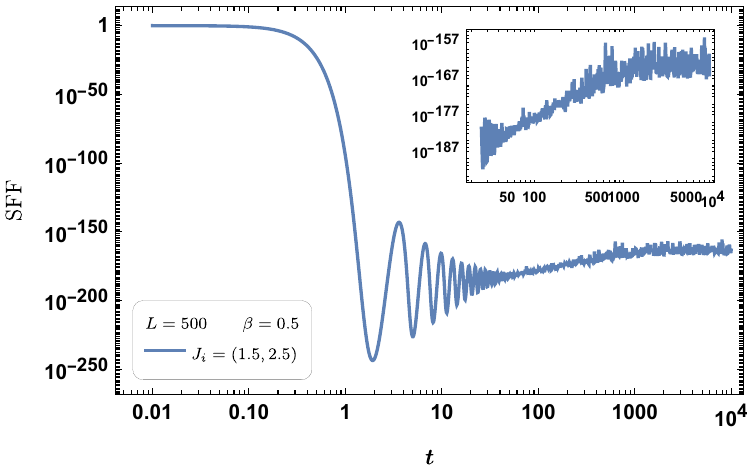}
\hspace{1mm}
\includegraphics[scale=0.4]{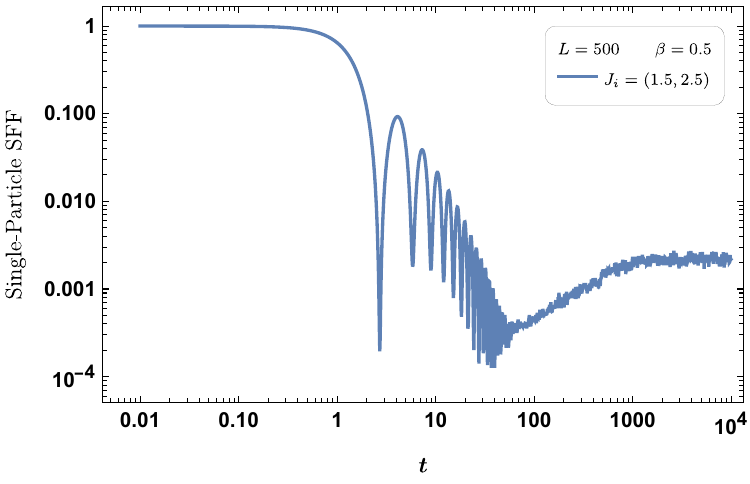}
\hspace{1mm}
\includegraphics[scale=0.4]{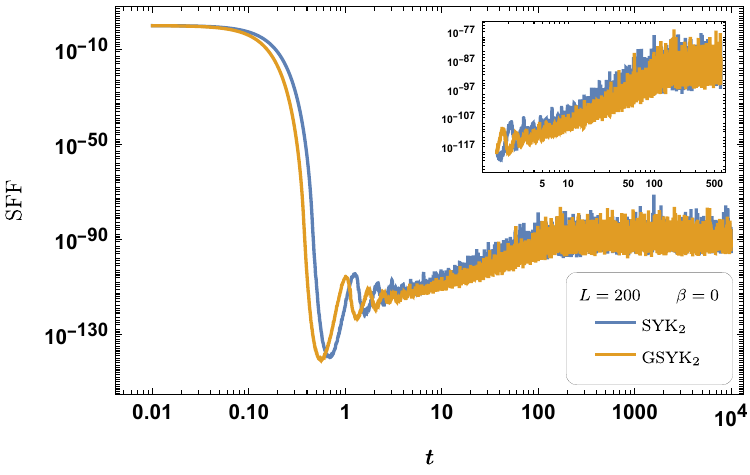}
\end{center}
\caption{This figure shows the spectral form factor for two cases: (i) the full spectrum of the model and (ii) the single-particle sector of the spectrum. The two plots on the left correspond to the local disordered Ising model. In the inset of the left plot, we have magnified the main plot to show the ramp clearly. The right plot corresponds to GSYK$_2$ versus SYK$_2$. In all plots, the results are averaged over 2000 samples.
}
\label{fig:SFF}
\end{figure*}

In Fig.~\ref{fig:I3}, we illustrate the behavior of the TMI in our random integrable fermionic models. We divide the total system into four adjacent pieces denoted by $A_i$ where $i=1,2,3,4$ and study $I(A_1:A_2:A_3)$. In all cases, the dynamics is studied after a quantum quench, with the initial (pre-quench) state, $\ket{\psi_0}$, chosen as the ground state of the Ising Hamiltonian with the transverse magnetic field shifted by one unit, i.e. $h_c+1$, where $h_c$ denotes the critical value of $h$. This initial state is evolved by our random Hamiltonians. The upper panel corresponds to local models. For the clean Ising chain (without randomness), the TMI initially takes negative values and later rises, exhibiting periodic revivals due to entanglement revivals in a finite system (see, e.g., \cite{PhysRevLett.112.220401}). In contrast, on an infinite lattice, the TMI saturates to zero. Here, however, we restrict ourselves to finite systems to make meaningful comparisons with the disordered models. For the random Ising chain, we observe that increasing the width of the coupling distribution $J_i$ gradually suppresses the oscillatory behavior, which eventually disappears for $J_i \in (1,3)$. Notably, the saturation value of $-I_3$ is larger for the narrower distribution $J_i \in (1.5,2.5)$ than for $J_i \in (1,3)$. This behavior can be understood in light of the results presented in Section~\ref{sec:level}, where we show that the level statistics ratio of the single-particle sector corresponding to $J_i \in (1.5,2.5)$ is closer to a chaotic level statistics ratio, compared to that of $J_i \in (1,3)$.

In the lower panel of Fig.~\ref{fig:I3}, we present the behavior of TMI in the nonlocal random models. For both particle-conserving and nonconserving random Hamiltonians, we observe qualitatively similar behavior; however, the nonconserving cases exhibit a larger lower bound for $I_3$. This lower bound can be interpreted as the most negative value that the TMI of a fermionic Gaussian state can attain under quadratic Hamiltonian evolution. More precisely, we consider the case where the fermionic operators are mixed by a random Haar unitary, $U c_j U^\dagger = \mathbf{U} c_j,$
with the matrix $\mathbf{U}$ drawn from the Haar measure over the unitary group. This corresponds to applying the transformation $U_{ik} \bra{\psi_0} c_k^\dagger c_l \ket{\psi_0} U_{lj},$ and analogous ones, to all blocks of the correlation matrix. The dashed line in the right panel represents the value of TMI for this Haar-randomly evolved state; it scales linearly with the system size and coincides with the saturation value of TMI in the GSYK$_2$ model.

It is well established that the average entanglement entropy of fermionic Gaussian states is substantially lower than that of generic (Haar-random) states. Although the average entropy of these states still obeys a volume law, the proportionality coefficient is smaller than that of generic typical states \cite{Vidmar:2017uux, PhysRevE.100.062134, Lydzba:2020qfx, Lydzba:2021hml, Bianchi:2021aui}. While, to the best of our knowledge, the average TMI for Gaussian random states has not been studied analytically, as our numerical results suggest, it would be natural to expect that the corresponding lower bound for the average TMI is also significantly less negative than that of Haar-random states, for which the lower bound at the leading order is given by $-2S + 1$, where $S$ denotes the average entropy of the smallest subsystem \cite{Hosur:2015ylk}. 

\begin{figure*}[t]
\begin{center}
\includegraphics[scale=0.26]{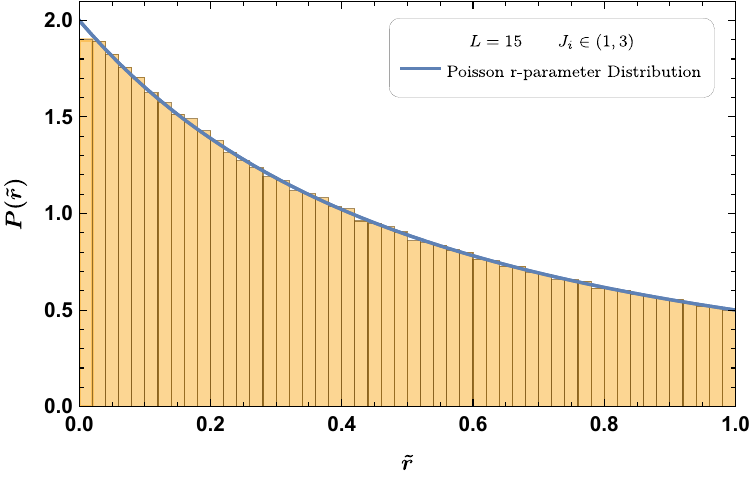}
\hspace{1mm}
\includegraphics[scale=0.26]{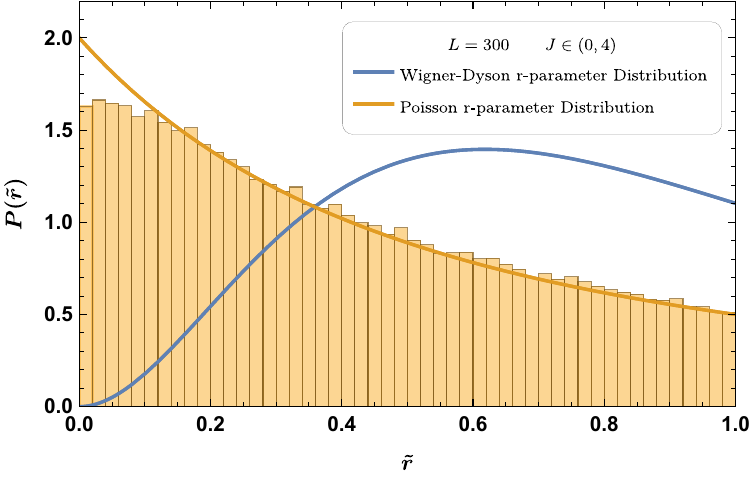}
\hspace{1mm}
\includegraphics[scale=0.26]{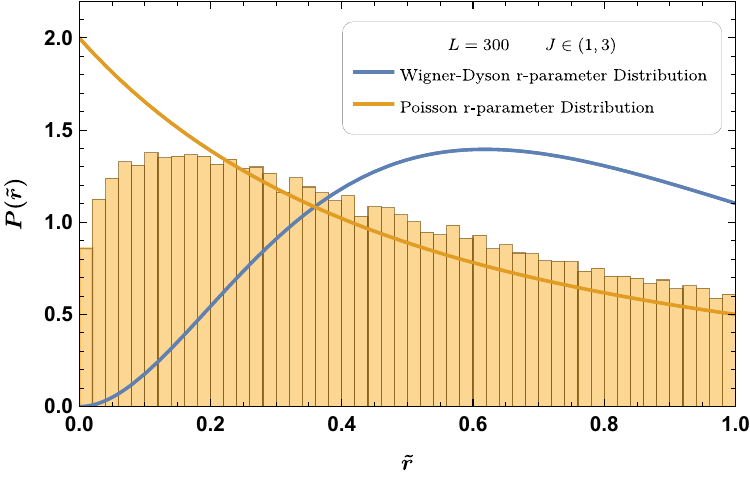}
\hspace{1mm}
\includegraphics[scale=0.26]{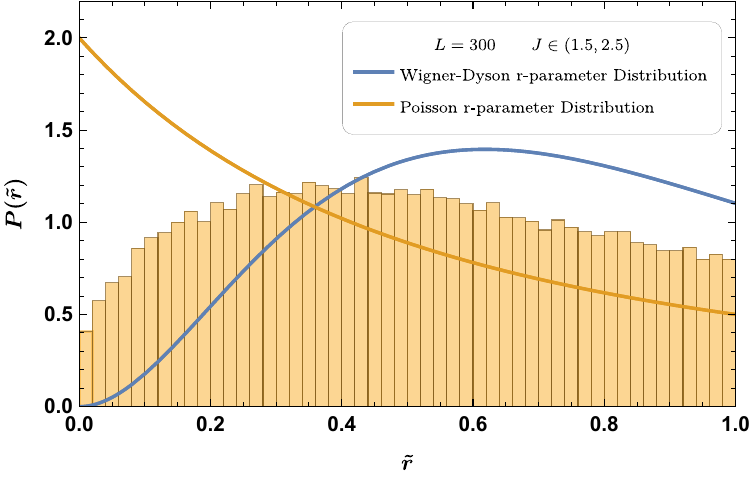}
\hspace{1mm}
\includegraphics[scale=0.26]{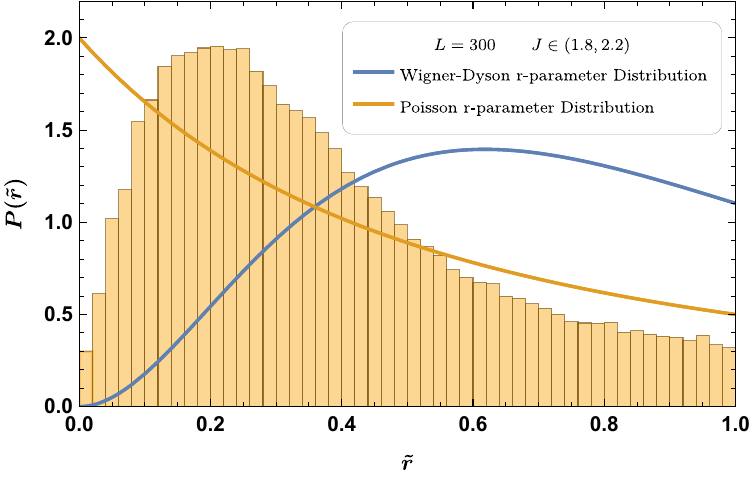}
\end{center}
\caption{The level-statistics ratio for the local random model. The rightmost panel shows the level-statistics ratio for the full spectrum, which follows a Poisson $r$-parameter distribution. Moving from right to left, starting from the second panel, the remaining panels display the level-statistics ratio for the single-particle sector, arranged in order of decreasing range of the random coupling. The corresponding coupling ranges are indicated in the legends of each panel. From left to write $\langle \tilde{r} \rangle= 0.386, 0.397, 0.431, 0.499, 0.373$. For reference, the Poisson and Wigner–Dyson distributions of the $r$-parameter are also shown for comparison. All results are averaged over 300 samples.
}
\label{fig:LevelStat}
\end{figure*}

\section{Spectral Form Factor}\label{sec:sff}
The spectral form factor has emerged as a central diagnostic in the study of quantum chaos and information scrambling \cite{Haake:2010fgh, Cotler:2016fpe}. Defined as the Fourier transform of the two-point correlation function of the energy spectrum, the SFF provides a direct window into the statistical properties of many-body spectra. In particular, its behavior at intermediate and late times distinguishes chaotic systems, characterized by universal random matrix features such as the \textit{linear} ramp and plateau, from integrable systems, which typically lack these signatures. Thus, the SFF serves as a powerful tool to bridge spectral statistics with dynamical indicators of scrambling.

In the present context of quadratic random dynamics, the spectral form factor (SFF) takes on particular significance. Although these systems are quantum integrable, the introduction of randomness alters their spectral correlations in nontrivial ways. Investigating the SFF in this setting therefore provides a means to probe how randomness alone—without interactions beyond the quadratic level—can induce spectral features reminiscent of chaos. In this sense, the SFF offers a spectral counterpart to entanglement-based diagnostics, enriching our understanding of scrambling in Gaussian systems. This perspective has been partially developed in earlier works: in \cite{Winer:2020mdc, Liao:2020lac} the SYK$_2$ model was shown to exhibit an exponential, rather than linear, ramp, while in \cite{Mollabashi:2024gik} random quadratic bosonic Hamiltonians were found to display a nonlinear ramp. Taken together, these observations suggest that the very presence of a ramp may serve as an indicator of scrambling, while chaotic systems can be distinguished by the \textit{linearity} of the ramp.

The SFF defined via the analytic continuation of the partition function $Z(\beta)$ as $g(\beta,t)=\frac{|Z(\beta+it)|^2}{|Z(\beta)|^2}$ can be expressed in terms of the decoupled modes  
\begin{equation}
    g(\beta,t)=\prod_k g_k(\beta,t)
\end{equation} 
where
\be
    g_k(\beta,t)=\frac{|Z_k(\beta+it)|^2}{|Z_k(\beta)|^2}=1+\frac{\cos ( \lambda_k t)-1}{\cosh (\beta  \lambda_k)+1}\,.
\ee

In Fig.~\ref{fig:SFF}, we present the numerical results for the spectral form factor (SFF). The left panel corresponds to our local random model, where, following the initial dip—with characteristic oscillations—a ramp develops before reaching the plateau. The inset highlights this ramp region more clearly. The middle panel shows the SFF for the single-particle sector of the local model. For a sufficiently broad range of the random coupling parameter $J_i$, the ramp exhibits an almost linear behavior. For the SYK$_2$ and GSYK$_2$ models, whose single-particle sectors inherently obey Wigner–Dyson statistics, the appearance of a linear ramp is expected. However, for the local random models, where the spectral statistics will be analyzed in the next section, the emergence of such a ramp is non-trivial. The right panel shows the SFF for SYK$_2$ and GSYK$_2$. While the analytic behavior of SYK$_2$ is well known, our numerical results demonstrate that GSYK$_2$ exhibits qualitatively similar behavior.

\section{Spectral Statistics}\label{sec:level}

The eigenvalues of chaotic Hamiltonians exhibit statistical properties analogous to those of random matrices, with maximally chaotic systems following the predictions of Random Matrix Theory (RMT)~\cite{PhysRevLett.52.1}. Denoting the ordered eigenvalues by $E_n$ ($E_{n+1} > E_n$), the level spacing $S_n = E_{n+1} - E_n$ serves as a key indicator of chaoticity: integrable systems display Poisson statistics, $P(s) = e^{-s}$, while maximally chaotic systems follow Wigner-Dyson statistics, $P(s) = A s^\beta e^{-B s^2}$, with $\beta = 1, 2, 4$ and normalization constants $A$ and $B$. Here $P(s)$ gives the probability density that two consecutive eigenvalues have spacing $s$. Since the raw level spacings depend on the local density of states and are therefore nonuniversal, one typically performs \emph{spectral unfolding}, a rescaling by the local mean level spacing, to reveal universal statistical features, although this procedure is inherently model-dependent and lacks a universally optimal definition.

To overcome the ambiguities of spectral unfolding, Ref.~\cite{PhysRevB.75.155111} introduced the distribution of ratios of consecutive level spacings, defined as
$$\tilde{r}_n = \frac{\min(s_n,s_{n-1})}{\max(s_n,s_{n-1})} = \min(r_n,1/r_n)$$ with $r_n = s_n/s_{n-1}$. This quantity provides a universal diagnostic independent of the local density of states. Ref.~\cite{PhysRevLett.110.084101} further derived analytic Wigner-like surmises for the three classical random-matrix ensembles, yielding the distribution $P_W(\tilde{r}) = Z_\beta^{-1} (\tilde{r}+\tilde{r}^2)^{\beta} / (1+\tilde{r}+\tilde{r}^2)^{1+\tfrac{3}{2}\beta}$, where $\beta$ is the Dyson index and $Z_\beta$ ensures normalization. The average value $\langle \tilde{r} \rangle$ provides a robust quantitative benchmark for spectral correlations. For uncorrelated (Poisson) spectra, the corresponding ratio distribution $P_{\text{Poisson}}(\tilde{r}) = 2/(1+\tilde{r})^2$ characterizes integrable systems. The benchmark values of $\langle \tilde{r} \rangle$ are $\langle \tilde{r} \rangle \approx 0.386$ for Poisson statistics, and $0.536$, $0.602$, and $0.676$ for the GOE, GUE, and GSE ensembles, respectively.
\begin{figure}[h]
\begin{center}
\includegraphics[scale=0.32]{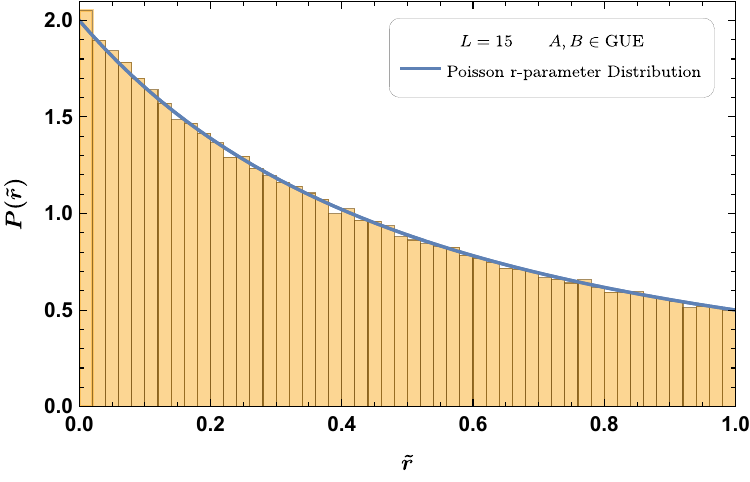}
\includegraphics[scale=0.32]{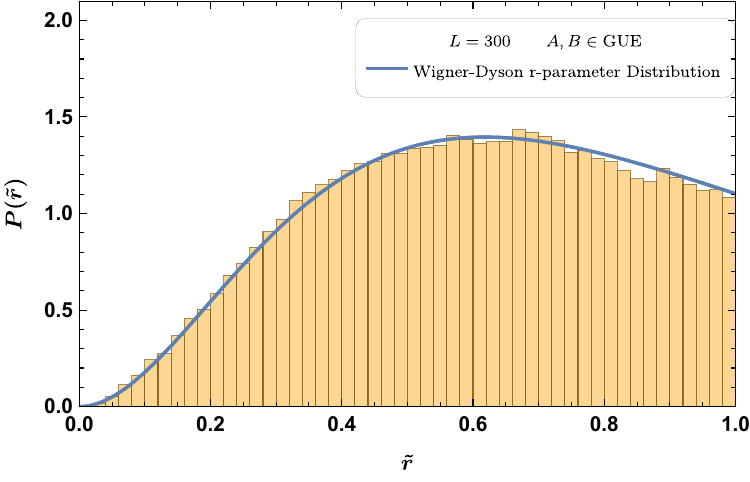}
\end{center}
\caption{The level-statistics ratio for GSKY$_2$ model. The left panel following a Poisson $r$-parametr distribution corresponds to the full spectrum, and the right panel following a Wigner-Dyson $r$-parametr distribution corresponds to the single-particle sector. For the left and right panels we find $\langle \tilde{r} \rangle \approx 0.384, 0.598$, respectively. The results are averaged over 300 samples.
}
\label{fig:LevelStatNL}
\end{figure}

In Fig.~\ref{fig:LevelStat}, we show the level-spacing ratio for the local random model. As expected from its integrable nature, the ratio for the full spectrum, displayed in the left panel, follows a Poisson-like distribution with $\langle \tilde{r} \rangle \approx 0.386$. Remarkably, in the single-particle sector of the local model, we observe a mixed behavior in the level-spacing ratio as the range of the random couplings $J_i$ is varied. From the second right panel to the left, the corresponding values of $\langle \tilde{r} \rangle$ are 0.397, 0.431, 0.499, and 0.373, for $J_i\in(0,4)$, $J_i\in(1,3)$, $J_i\in(1.5,2.5)$, and $J_i\in(1.8,2.2)$, respectively. In other words, for sufficiently broad coupling ranges, $\langle \tilde{r} \rangle$ starts close to the Poisson value and gradually increases toward a Wigner-Dyson-like value as the range narrows. Interestingly, when the coupling distribution becomes extremely narrow, for instance, $J_i\in(1.8,2.2)$, the average ratio $\langle \tilde{r} \rangle$ deviates from this trend and no longer lies between the Poisson-like and Wigner-Dyson-like limits.

In Fig.~\ref{fig:LevelStatNL}, we show the level-statistics ratio corresponding to the nonlocal GSKY$_2$ model. The behavior for the full level statistics nicely follows a Poisson-like distribution with $\langle \tilde{r} \rangle \approx 0.384$ and the single particle sector nicely follows a Wigner–Dyson-like distribution with $\langle \tilde{r} \rangle \approx 0.598$. 

\section{conclusions and discussions}\label{sec:conclusion}
In this work, we investigated the role of randomness in the scrambling of quantum information within integrable free-fermionic models, demonstrating that randomness, serving as a minimal ingredient for scrambling, can induce information scrambling even in integrable systems. Specifically, we considered quadratic Hamiltonians with varying degrees of randomness and analyzed several measures to characterize the resulting scrambling behavior. We showed that the memory effect in the entanglement of disjoint blocks in a Gaussian state disappears when the local random couplings are drawn from a sufficiently broad distribution. This disappearance corresponds to the delocalization of information as quantified by the mutual information. The behavior of the tripartite mutual information (TMI) is qualitatively similar to that of chaotic systems, in that it takes negative values and saturates at a negative constant. However, the lower bound of the TMI in these integrable quadratic systems is significantly smaller in magnitude than in chaotic systems, indicating that scrambling in Gaussian states governed by integrable quadratic Hamiltonians is substantially weaker than in generic states evolved under chaotic dynamics.

It is worth emphasizing that, since TMI probes genuine multipartite entanglement beyond bipartite correlations, the standard quasiparticle picture is not expected to faithfully capture its dynamics in our systems of interest. In particular, the conventional quasiparticle framework predicts that TMI should vanish whenever at least two of the subsystems are adjacent \cite{Parez:2022egh}, whereas several studies have reported that TMI can display nontrivial temporal behavior even in free systems. A refined quasiparticle description that accounts for the existence of multiplets comprising more than two mutually entangled quasiparticles \cite{Bertini:2018ymp, Bastianello:2018fvl, Roy:2021efl, Carollo:2022lrq, Caceffo:2023hns} is therefore expected to more accurately describe the behavior of TMI in our disordered models. A detailed analysis of this perspective in random quadratic systems is left for future work.

We also examined the spectral properties of these systems by studying the spectral form factor (SFF) and level-spacing statistics. Our numerical results indicate that the SFF of local random models exhibits a ramp structure, despite the absence of any (completely) chaotic sector. Interestingly, this ramp structure of GSYK$_2$ closely resembles that of SYK$_2$ model.

Furthermore, we analyzed the level-spacing ratio in these models. While the full spectrum follows a Poisson-like distribution, the single-particle sector exhibits a notable transition: as the range of the local random couplings is narrowed, the level-spacing ratio evolves from a Poisson-like toward a Wigner–Dyson-like distribution. This transition, however, remains only partial, when the coupling range becomes extremely narrow, the level-spacing ratio deviates from this trend. Although a comprehensive analysis of the ultra-narrow coupling regime is left for future work, this partial transition provides valuable insight: the smallest saturation value of the TMI, coinciding with that of the SYK$_2$ model, occurs precisely for the coupling range that yields the largest average level-spacing ratio, i.e., the value closest to that of chaotic systems.

Finally, we note that while the behavior of out-of-time-ordered correlators (OTOCs), a well-established probe of quantum information scrambling, has been examined for specific operators in the integrable SYK$_2$ model \cite{Garcia-Garcia:2024mdv}, a more comprehensive analysis of OTOCs in random integrable systems, particularly in local random settings and for operators restricted to the single-particle sector, would be an interesting direction for future work.

\subsection*{Acknowledgments} We thank Saleh Rahimi-Keshari for valuable discussions and Gilles Parez for fruitful comments on the first version of this work. MV is grateful to Mohsen Alishahiha for fruitful discussions and support. AM would like to acknowledge support from the Iran National Elites Foundation under the Kazemi–Ashtiani starting grant. MV is supported by Iran National Science Foundation (INSF)
under project No.4023620. 

\bibliography{bib}

\end{document}